\documentclass[twoside]{article}
\usepackage{qic,epsfig,bbm}

\begin{document}
\setlength{\textheight}{8.0truein}    

\runninghead{Phase estimation using an approximate eigenstate}
            {Avatar Tulsi }

\normalsize\textlineskip
\thispagestyle{empty}
\setcounter{page}{1}

\copyrightheading{0}{0}{2003}{000--000}

\vspace*{0.88truein}

\alphfootnote

\fpage{1}

\centerline{\bf PHASE ESTIMATION USING AN APPROXIMATE EIGENSTATE}
\vspace*{0.37truein}
\centerline{\footnotesize AVATAR TULSI}
\vspace*{0.015truein}
\centerline{\footnotesize\it Department of Physics, IIT Bombay}
\baselineskip=10pt
\centerline{\footnotesize\it Mumbai, 400076, India}
\vspace*{0.225truein}
\publisher{(received date)}{(revised date)}

\vspace*{0.21truein}

\abstracts{A basic building block of many quantum algorithms is the Phase Estimation algorithm (PEA). It estimates an eigenphase $\phi$ of a unitary operator $U$ using a copy of the corresponding eigenstate $|\phi\rangle$. Suppose, in place of $|\phi\rangle$, we have a copy of an approximate eigenstate $|\psi\rangle$ whose overlap magnitude with $|\phi\rangle$ is at least $\sqrt{2/3}$. Then PEA fails with a constant probability. However, using multiple copies of $|\psi\rangle$, the failure probaility can be made to decrease exponentially with the number of copies. In this paper, we show that as long as we can perform a selective inversion of $|\psi\rangle$, a single copy is sufficient to estimate $\phi$.} {An important application is to improve the spatial complexity of eigenpath traversal algorithm, a "digital" analogue of quantum adiabatic evolution, having applications ranging from quantum physics simulation to optimization. Here the goal is to travel a path of eigenstates of $n$ different unitary operators satisfying some conditions. The fastest algorithm is due to Boixo, Knill and Somma (BKS) which needs $\Theta(\ln n)$ copies of the eigenstate. Using our algorithm, BKS algorithm can work using just a single copy of the eigenstate.}{}

\vspace*{10pt}

\vspace*{3pt}
\communicate{to be filled by the Editorial}

\vspace*{1pt}\textlineskip    

\section{Introduction}	       
\vspace*{-0.5pt}
\noindent
Phase estimation algorithm (PEA) is the backbone of many important quantum algorithms~\cite{PEA,CEMM,phase,shor}. Let $|\phi\rangle$ be an eigenstate of a unitary operator $U$ with the eigenvalue $e^{i 2\pi \phi}$. PEA attaches $\mu$ ancilla qubits to $|\phi\rangle$ and starts with the state $|\phi,0_{\mu}\rangle = |\phi\rangle|0_{\mu}\rangle$, where $|0_{\mu}\rangle$ denotes the state of all $\mu$ ancilla qubits in $|0\rangle$ state. PEA applies the operator $\mathcal{E}(U)$ on $|\phi\rangle|0_{\mu}\rangle$ to get the state $|\phi,f\rangle = |\phi\rangle|f(\phi)\rangle$. The $|f(\phi)\rangle$ state provides an estimate of $\phi$ as its amplitudes are mostly within a narrow interval of basis states $|k\rangle$'s of $\mu$ ancilla qubits. The interval is centred at $|k(\phi)\rangle$, the basis state encoding $k(\phi) = \lfloor 2^{\mu}\phi \rfloor$. The accuracy of estimation increases exponentially with $\mu$ and the PEA operator $\mathcal{E}(U)$ uses $2^{\mu}$ applications of $U$.

	Suppose we don't have a copy of the perfect eigenstate $|\phi\rangle$.  Rather what we have is a copy of an approximate eigenstate $|\psi\rangle$ satisfying $|\langle \psi|\phi\rangle| = \alpha  \geq \sqrt{2/3}$. We expand $|\psi\rangle$ in the eigenbasis of $U$ as $|\psi\rangle = \alpha |\phi\rangle  + \sum_{d}\alpha_{d}|\phi^{\perp d}\rangle$, where $|\phi^{\perp d}\rangle$ denotes non-$|\phi\rangle$ eigenstates of $U$. Applying $\mathcal{E}(U)$ on $|\psi,0_{\mu}\rangle$, we get
\noindent   
\begin{equation}
\mathcal{E}(U)|\psi,0_{\mu}\rangle = |\psi,f\rangle = \alpha |\phi\rangle|f(\phi)\rangle + \sum_{d}\alpha_{d}|\phi^{\perp d}\rangle |f(\phi^{\perp d})\rangle. \label{PEA2}
\end{equation}
The component of above state in $|f(\phi)\rangle$ state is $\alpha$ and we get an estimate of $\phi$ with probability $\alpha^{2} \geq 2/3$. So the failure probability can be as high as $1/3$.

	If $|\psi\rangle$ is easy to prepare then we just do $r$ preparations of $|\psi\rangle$, subsequently apply $\mathcal{E}(U)$, and then measure ancilla qubits. The failure probability decreases exponentially with $r$. However, in many situations of interest, $|\psi\rangle$ is not so easy to prepare. For example, if $|\psi\rangle$ is the output of a quantum algorithm then the algorithm must be applied every time to prepare a fresh copy of $|\psi\rangle$. In such cases, it becomes crucial to keep $|\psi\rangle$ state intact. Multiple copies of $|\psi\rangle$ can reduce the failure probability. Let $|\psi\rangle^{\otimes r}$ denote the joint state of $r$ copies of $|\psi\rangle$. With each copy, we attach a register of $\mu$ ancilla qubits. We do a parallel application of $\mathcal{E}(U)$ to get $\left[\mathcal{E}(U)|\psi,0_{\mu}\rangle\right]^{\otimes r}$. Eq. (\ref{PEA2}) implies  
\noindent   
\begin{equation}
\left[\mathcal{E}(U)|\psi\rangle|0_{\mu}\rangle\right]^{\otimes r} = \chi_{>}|>\rangle + \chi_{<}|<\rangle.
 \label{multiplePEAmore}
\end{equation}
Here $|>\rangle$ ($|<\rangle$) denotes a normalized quantum state in which more (less) than $r/2$ registers are in $|f(\phi)\rangle$ state. 

	It is easy to do a projective measurement $\mathcal{M}_{>,<}$ in the basis $\{|>\rangle, |<\rangle\}$ by distinguishing between $|>\rangle$ and $|<\rangle$ states. We can reversibly compute if more than half registers are in $|f(\phi)\rangle$ state as the amplitudes of $|f(\phi)\rangle$ are mostly within a narrow interval of $|k\rangle$'s. This distinguishes $|>\rangle$ from $|<\rangle$ and implements $\mathcal{M}_{>,<}$.	After the measurement $\mathcal{M}_{>,<}$ of $\left[\mathcal{E}(U)|\psi,0_{\mu}\rangle\right]^{\otimes r}$, we get $|>\rangle$ state with probability $1-\chi_{<}^{2}$. The failure probability is $\chi_{<}^{2}$. Using $|>\rangle$ state, we can reversibly compute $k(\phi)$ to estimate $\phi$. Then we apply $\left[\mathcal{E}^{\dagger}(U)\right]^{\otimes r}$ on $|>\rangle$ state to get $\left(|\psi,0_{\mu}\rangle\right)^{\otimes r} + |\chi_{<}\rangle$, where $|\chi_{<}\rangle$ is a state of length $\chi_{<}$. Thus, as desired, all $r$ copies of eigenstates remain intact with an error term $O(\chi_{<})$. 

	To upper bound the error term $O(\chi_{<})$, note that the probability of finding a single register in $|f(\phi)\rangle$ state is $\alpha^{2} \geq 2/3$. Hence the expected number of registers in $|f(\phi)\rangle$ state is at least $2r/3$. Due to Hoeffding's bound (~\cite{hoeffding}), the probability of getting less than $r/2$ registers in $|f(\phi)\rangle$ is at most $e^{-2rt^{2}}$ where $t = (2/3)-(1/2) = 1/6$. Thus $\chi_{<}^{2} \leq e^{-r/18}$ or $\chi_{<} \leq e^{-r/36} = e^{-\Theta(r)}$. Suppose a quantum algorithm uses $n$ instances of phase estimation with approximate eigenstates. Then the error in each instance must be less than $1/n$ for a successful algorithm. So we must have $r = \Theta(\ln n)$ copies of $|\psi\rangle$. 

	In this paper, we show that as long as we can implement a selective inversion $I_{\psi}$ of $|\psi\rangle$, a single copy is sufficient to estimate $\phi$. Implementation of $I_{\psi}$ is possible in some cases, for example, if $|\psi\rangle$ is an eigenstate of a known operator with a known eigenvalue. As an application, we discuss the eigenpath traversal problem. Let $|\theta_{s}\rangle$ be the non-degenerate eigenstates of unitary operators $V_{s}$ for $s \in \{1,2,\ldots,n\}$. The goal is to evolve a copy of $|\theta_{1}\rangle$ to $|\theta_{n}\rangle$. This problem has applications ranging from quantum physics simulation to optimization. The algorithm with optimal time complexity is due to Boixo, Knill and Somma (BKS)~\cite{BKS}. BKS algorithm uses $n$ instances of PEA with approximate eigenstates and hence it needs $\Theta(\ln n)$ copies of $|\theta_{1}\rangle$. Using our algorithm, BKS algorithm works with just a single copy. Hence we get a significant improvement in the spatial complexity of BKS algorithm. 

	We present the main algorithm in the next section. The probability estimation algorithm, used as a subroutine, is presented in Section III. We discuss the application to eigenpath traversal problem in Section IV. We discuss and conclude in Section V.

\section{Phase Estimation with approximate eigenstate}
\noindent
In our algorithm, we will be using PEA for unitary operators $U_{\gamma}$ having $|\phi\rangle$ as its eigenstate with the eigenvalue $e^{i 2\pi \phi_{\gamma}}$ given by  
\noindent   
\begin{equation}
U_{\gamma} = U^{2^{\gamma}} \Longrightarrow U_{\gamma}|\phi\rangle = e^{i 2\pi \phi_{\gamma}}|\phi\rangle,\ \ \phi_{\gamma} = 2^{\gamma}\phi. 
\end{equation}
Note that $U_{0} = U$ and $\phi_{0} = \phi$. We choose $\mu = 5$ ancilla qubits in our usage of PEA. We work in the joint Hilbert space, $\mathcal{H}_{J} = \mathcal{H}_{m} \otimes \mathcal{H}_{32}$, where $\mathcal{H}_{m}$ is the Hilbert space of main register spanned by all possible eigenstates $|\phi\rangle$ of $U$ and $\mathcal{H}_{32}$ is the $32$-dimensional Hilbert space of $5$ ancilla qubits. Each basis state $|k\rangle$ of $\mathcal{H}_{32}$ encode an integer $k \in \{0,1,\ldots,31\}$, the decimal value of the binary number encoded by $|k\rangle$. 

The PEA operator $\mathcal{E}(U_{\gamma})$ acts on $\mathcal{H}_{J}$ as
\begin{equation}
\mathcal{E}(U_{\gamma})|\phi,0_{\mu}\rangle = |\phi,f_{\gamma}\rangle = |\phi\rangle|f(\phi_{\gamma})\rangle.
\end{equation}
With approximate eigenstate $|\psi,0_{\mu}\rangle$, the action of $\mathcal{E}(U_{\gamma})$ can be written using Eq. (\ref{PEA2}) as
\begin{equation}
\mathcal{E}(U_{\gamma})|\psi,0_{\mu}\rangle = |\psi,f_{\gamma}\rangle = \alpha|\phi\rangle|f(\phi_{\gamma})\rangle + \sum_{d}\alpha_{d}|\phi^{\perp d}\rangle|f(\phi_{\gamma}^{\perp d})\rangle. \label{PEA3}
\end{equation}
The $|f(\phi_{\gamma})\rangle$ state provides an estimate of $\phi_{\gamma}$. To quantify this estimate, we use the results of Sec. 5.2.1 of ~\cite{phase}, which is based on the analysis of ~\cite{CEMM}. Let $[a,b]$ denote the set of integers ranging from $a$ $\rm mod 32$ to $b$ $\rm mod 32$ in the increasing order. In modular arithmetic, the increasing order is defined even if $b < a$. For example, $[16,15]$ is the set $\{16,17,\ldots,31,0,\ldots,15\}$ in the increasing order, not $\{16,15\}$. Let $\Lambda_{a}^{b}$ denote the subspace of $\mathcal{H}_{J}$ in which $|k\rangle$, the basis states of ancilla qubits, satisfy $k \in [a,b]$ and let $\Pi_{a}^{b}$ denote the projection operator on $\Lambda_{a}^{b}$. For any state $|\sigma\rangle$ in $\mathcal{H}_{J}$, the probability $P_{a}^{b}(\sigma)$ of getting an integer $k \in [a,b]$ after measuring the ancilla qubits is
\noindent   
\begin{equation}
P_{a}^{b}(\sigma) = |\langle \sigma|\lambda_{a}^{b}(\sigma)\rangle|^{2},\ \ \ |\lambda_{a}^{b}(\sigma)\rangle = \Pi_{a}^{b}|\sigma\rangle.  \label{probabilityabdefined}
\end{equation}

	Eq. (5.34) of Sec. 5.2.1 of ~\cite{phase} can be rewritten in our notation as
\noindent   
\begin{equation}
P_{k(\phi_{\gamma})-c}^{k(\phi_{\gamma})+c}(\phi,f_{\gamma}) \geq 1-\frac{1}{2(c-1)},\ \ \ \ k(\phi_{\gamma}) = \lfloor 2^{\mu}\phi_{\gamma} \rfloor,\ \ \ c > 1. \label{probabilityeqn}
\end{equation}
We choose $c = 4$. Then
\noindent   
\begin{equation}
P_{k(\phi_{\gamma})\pm 4}(\phi,f_{\gamma}) = P_{k(\phi_{\gamma})-4}^{k(\phi_{\gamma})+4}(\phi,f_{\gamma}) \geq 5/6. \label{eqnc4}
\end{equation}
This also gives a lower bound on $P_{k(\phi_{\gamma})\pm 4}(\psi,f_{\gamma})$ where $|\psi,f_{\gamma}\rangle$, the output state of PEA with approximate eigenstate $|\psi\rangle$, is given by Eq. (\ref{PEA3}). The eigenstates $|\phi^{\perp d}\rangle$ are orthogonal to $|\phi\rangle$ and hence the corresponding states $|\phi^{\perp d}\rangle|f(\phi_{\gamma}^{\perp d})\rangle$ only add to the measurement probabilities of getting any $k$. Thus
\noindent   
\begin{equation}
P_{k(\phi_{\gamma})\pm 4}(\psi,f_{\gamma}) \geq \alpha^{2}P_{k(\phi_{\gamma})\pm 4}(\phi,f_{\gamma}) \geq (2/3) \times (5/6) = (5/9)\,  \label{lowerbound1}
\end{equation} 
which is the desired lower bound.	

	This bound helps in estimating $\phi_{\gamma}$. Let $\rm MSB_{\gamma}[x]$ denote the $(\gamma + 1)^{\rm th}$ most significant bit of its argument $x$. Thus $\gamma$ can be zero or any positive integer. We can use above bound in finding $\rm MSB_{\gamma}[\phi]$ for any $\gamma$. Suppose $k(\phi_{\gamma}) = \lfloor 32\phi_{\gamma}\rfloor \in [0,7]$. Then 
\noindent   
\begin{equation}
k(\phi_{\gamma}) \in [0,7] \Longrightarrow [k(\phi_{\gamma}) \pm 4] = [k(\phi_{\gamma})-4,k(\phi_{\gamma})+4] \subset [28,11].
\end{equation} 
By definition of $[a,b]$ and $P_{a}^{b}(\sigma)$, $[a,b] \subset[a',b']$ implies that $P_{a'}^{b'}(\sigma) \geq P_{a}^{b}(\sigma)$. Using Eq. (\ref{lowerbound1}), we get
\noindent   
\begin{equation}
k(\phi_{\gamma}) \in [0,7] \Longrightarrow P_{28}^{11}(\psi,f_{\gamma}) \geq 5/9,\ \ \ P_{28}^{11}(\psi,f_{\gamma}) < 5/9 \Longrightarrow k(\phi_{\gamma}) \notin [0,7]. \label{imply1} 
\end{equation}
A mere relabeling of the basis states as $|k\rangle \longrightarrow |(k+16)\rm mod 32\rangle$ does not affect above relation and we get $P_{12}^{27}(\psi,f_{\gamma}) < 5/9 \Longrightarrow k(\phi_{\gamma}) \notin [16,23]$. By definition, $P_{12}^{27}(\sigma)$ and $P_{28}^{11}(\sigma)$ are interrelated as $P_{12}^{27}(\sigma)+P_{28}^{11}(\sigma) = 1$. Hence we get
\noindent   
\begin{equation}
P_{12}^{27}(\psi,f_{\gamma}) \geq 4/9 \Longrightarrow k(\phi_{\gamma}) \notin [0,7],\ \ \ P_{12}^{27}(\psi,f_{\gamma}) < 5/9 \Longrightarrow k(\phi_{\gamma}) \notin [16,23]. \label{tableimply1}
\end{equation}
Relabeling the states again as $|k\rangle \longrightarrow |(k+8)\rm mod 32\rangle$, we get 
\noindent   
\begin{equation}
P_{20}^{3}(\psi,f_{\gamma}) \geq 4/9 \Longrightarrow k(\phi_{\gamma}) \notin [8,15],\ \ \ P_{20}^{3}(\psi,f_{\gamma}) < 5/9 \Longrightarrow k(\phi_{\gamma}) \notin [24,31].  \label{tableimply2}
\end{equation}
We define the notation
\begin{equation}
P_{0\gamma} = P_{12}^{27}(\psi,f_{\gamma}),\ \ \ \ \ P_{1\gamma} = P_{20}^{3}(\psi,f_{\gamma}).  
\end{equation}
In this notation, the relations (\ref{tableimply1}) and (\ref{tableimply2}) are tabulated in Table \ref{Table 1} in the columns $I$ to $VI$. The entries of Column $VII$ are easy to check. For example, if $k(\phi_{\gamma}) \notin [0,7]$ and $k(\phi_{\gamma}) \notin [24,31]$, the only possibility is $k(\phi_{\gamma}) \in [8,23]$ which is $k(\phi_{\gamma}) \in [01000,10111]$ in binary representation. Then $\rm MSB_{0}[k(\phi_{\gamma})] = 1-\rm MSB_{1}[k(\phi_{\gamma})]$. The entries of Column $VIII$ is due to the following facts, 
\begin{equation}
k(\phi_{\gamma}) = \lfloor 32 \phi_{\gamma} \rfloor \Longrightarrow \rm MSB_{\gamma}[k(\phi_{\gamma})] = \rm MSB_{\gamma}[\phi_{\gamma}],\ \ \rm for \gamma \in \{0,1\},  
\end{equation} 
\begin{equation}
\phi_{\gamma} = 2^{\gamma}\phi \Longrightarrow \rm MSB_{\gamma'}[\phi_{\gamma}] = \rm MSB_{\gamma  +\gamma'}[\phi].
\end{equation}

\begin{table}[h]
\tcaption{Finding $\rm MSB_{\gamma}[\phi]$ using the values of $P_{0\gamma}$ and $P_{1\gamma}$.}
\begin{tabular}{|c|c|c|c|c|c|c|c|}
\hline
I & II & III & IV & V & VI & VII & VIII \\
\hline
Case & $P_{0\gamma}$ & $k(\phi_{\gamma}) \notin $ & $ P_{1\gamma} $ & $ k(\phi_{\gamma}) \notin $ & $ k(\phi_{\gamma}) \in $ & $ \rm MSB_{0}[k(\phi_{\gamma})] $ &  $ \rm MSB_{\gamma}[\phi] $\\ 
\hline
 $1$ & $ \geq 4/9 $ & [0,7] & $ \geq 4/9 $ & $[8,15]$ &  $[16,31]$ & $ 1 $ & $ 1 $ \\
\hline
 $2$ & $ \geq 4/9 $ & $ [0,7] $ & $ <5/9 $ & $ [24,31] $ &  $[8,23]$ & $ 1-\rm MSB_{1}[k(\phi_{\gamma})] $ & $ 1-\rm MSB_{\gamma+1}[\phi] $\\
\hline
 $3$ & $ < 5/9 $ & $[16,23]$ & $\geq 4/9$ & $ [8,15]$ & $ [24,7] $ & $ \rm MSB_{1}[k(\phi_{\gamma})] $ & $ \rm MSB_{\gamma+1}[\phi] $ \\
\hline
 $4$ &  $< 5/9$ & $ [16,23] $ & $ < 5/9 $ & $ [24,31] $ & $ [0,15] $ &  $ 0 $ & $ 0 $ \\
\hline     
\end{tabular}
\label{Table 1}
\end{table}

In the next section, we present a probability estimation algorithm $\mathcal{S}_{g\gamma}$ to determine if $P_{g\gamma} \geq 4/9$ or $P_{g\gamma} < 5/9$ for $g = (0,1)$. Using Table \ref{Table 1}, either we find $\rm MSB_{\gamma}[\phi]$ in cases $(1,4)$ or we find it in terms of $\rm MSB_{\gamma + 1}[\phi]$ in cases $(2,3)$. We start with $\gamma =0$ and apply the algorithm $\mathcal{S}_{g0}$ for $g = (0,1)$. Either we find $\rm MSB_{0}[\phi]$ or we find it in terms of $\rm MSB_{1}[\phi]$. Then we increase $\gamma$ by $1$ so that $\gamma =1$. We apply the algorithm $\mathcal{S}_{g1}$ for $g = (0,1)$. Either we find $\rm MSB_{1}[\phi]$, which also allows us to find $\rm MSB_{0}[\phi]$, or we find it in terms of $\rm MSB_{2}[\phi]$. We keep on increasing $\gamma$ by $1$ and using the algorithm $\mathcal{S}_{g\gamma}$ for $g = (0,1)$. For $\gamma =\Gamma$, either we find $\rm MSB_{\Gamma}[\phi]$ in cases $(1,4)$, which also allows us to find $\rm MSB_{\gamma}[\phi]$ for all $\gamma < \Gamma$, or we find it in terms of $\rm MSB_{\Gamma +1}[\phi]$ in cases $(2,3)$. Suppose case $2$ is true. Then $k(\phi_{\Gamma}) \in [8,23]$. We define an operator $U^{+}$ as
\noindent 
\begin{equation}
U^{+}|\phi\rangle = e^{i 2\pi 2^{-\Gamma -2}}U|\phi\rangle = e^{i 2\pi \phi^{+}}|\phi\rangle,\ \ \ \ \phi^{+} = \phi + 2^{-\Gamma -2}. \label{uprimeequation} 
\end{equation}
Thus $\phi^{+}_{\Gamma} = 2^{\Gamma}\phi^{+} = \phi_{\Gamma} + 0.25$ and $k(\phi^{+}_{\Gamma}) = \lfloor 32\phi^{+}_{\Gamma}\rfloor = k(\phi_{\Gamma}) + 8$. As $k(\phi_{\Gamma}) \in [8,23]$ in case $2$, we get $k(\phi^{+}_{\Gamma}) \in [16,31]$ and hence $\rm MSB_{0}[k(\phi^{+}_{\Gamma})] = \rm MSB_{\Gamma}[\phi^{+}] = 1$. Similarly, if Case $3$ is true then $\rm MSB_{\Gamma}[\phi^{+}] = 0$. Thus we find $\rm MSB_{\Gamma}[\phi^{+}]$ in case we do not find $\rm MSB_{\Gamma}[\phi]$. We apply another round of the algorithm $\mathcal{S}_{g\gamma}$ for $g = (0,1)$ and for all $\gamma \leq \Gamma$ but this time we replace $U$ by $U^{+}$. Doing so will either give definite values of $\rm MSB_{\gamma}[\phi^{+}]$ for all $\gamma < \Gamma$ or their values in terms of $\rm MSB_{\Gamma}[\phi^{+}]$. Using the already known value of $\rm MSB_{\Gamma}[\phi^{+}]$, we determine $\rm MSB_{\gamma}[\phi^{+}]$ for all $\gamma \leq \Gamma$. 

	Thus we can find upto $(\Gamma + 1)^{\rm th}$ most significant bits of either $\phi$ or $\phi^{+}$. This determines either $\phi$ or $\phi^{+}$ to an accuracy of $2^{-\Gamma -1}$. As $\phi^{+} = \phi + 2^{-\Gamma -2}$, we find $\phi$ upto an accuracy of $2^{-\Gamma}$. For this, we need to apply the algorithm $\mathcal{S}_{g\gamma}$ for $g = (0,1)$ and for $\gamma \in \{0,1,\ldots, \Gamma\}$, once for $U$ and once for $U^{+}$, making a total of $4(\Gamma +1)$ applications. Thus if we want to estimate $\phi$ to an accuracy of $\delta = 2^{-\Gamma}$, we need
\begin{equation}
4[\log_{2}(1/\delta) + 1] \label{numberforphase}
\end{equation} 
applications of the algorithm $\mathcal{S}$ with $4$ applications for each value of $\gamma \in \{0,1,\ldots, \Gamma\}$. In next section, we describe the algorithm $\mathcal{S}$.

\section{Probability Estimation}
\noindent
By definition, $P_{a}^{b}(\sigma) = |\langle \sigma |\lambda_{a}^{b}(\sigma)\rangle|^{2}$ according to Eq. (\ref{probabilityabdefined}). Hence, in principle, it can be estimated by estimating the amplitude $|\langle \sigma|\lambda_{a}^{b}(\sigma)\rangle|$ using techniques based on standard quantum amplitude estimation~\cite{qaa2}. However the main purpose of those techniques is to estimate the amplitude, not to preserve the $|\sigma\rangle$ state, which is crucial for our purpose. For example, the overlap detection oracle used by BKS algorithm (Definition III.4 of ~\cite{BKS}) takes $O(1/\eta)$ time steps to output $1$ if the amplitude is less than $\eta_{0} -\eta$ and $0$ if it is more than $\eta_{0} + \eta$. For these cases, the algorithm can get back the $|\sigma\rangle$ state but when amplitude is in the interval $[\eta_{0}-\eta,\eta_{0}+\eta]$, it fails to do so. Choosing $\eta$ to be small increases the time complexity $O(1/\eta)$ and a careful calculation will show that the time complexity becomes significantly large in case when large number of estimations are needed. 

	We need a probability estimation algorithm which preserves the $|\sigma\rangle$ state with a sufficiently high probability. We present such an algorithm $\mathcal{S}$ in a general setting. Let the state $|\sigma\rangle$ has the projections $|\lambda\rangle$ and $|\lambda^{\perp}\rangle$ on two mutually complementary subspaces $\Lambda$ and $\Lambda^{\perp}$. Let $P_{\lambda}(\sigma)$ be the probability of getting $|\lambda\rangle$ state after measuring $|\sigma\rangle$. We have
\noindent   
\begin{equation}
|\sigma\rangle = \sin \omega |\lambda\rangle + \cos \omega |\lambda^{\perp}\rangle,\ \ \ P_{\lambda}(\sigma) = \sin^{2}\omega. \label{sigmalambda}
\end{equation}
Our goal is to determine if $P_{\lambda}(\sigma) \geq 4/9$ or $P_{\lambda}(\sigma) < 5/9$. Both of these inequalities may be true, but determining only one is sufficient for our purpose.

	 Though not necessary, but the algorithm becomes convenient if we can have a lower bound on $P_{\lambda}(\sigma)$. To do so, we attach an ancilla qubit in the state $(1/\sqrt{10})(|0\rangle + 3|1\rangle)$. The joint state $|\sigma'\rangle$ is given by
\noindent
\begin{equation}
\sqrt{10}|\sigma'\rangle = \sin \omega |0\rangle|\lambda\rangle + \cos \omega |0\rangle|\lambda^{\perp}\rangle + 3\sin \omega |1\rangle|\lambda\rangle + 3\cos\omega |1\rangle |\lambda^{\perp}\rangle \label{sigmaprimedefine}
\end{equation}
Let $\Lambda'$ be the subspace spanned by $\{|0\rangle|\lambda\rangle, |0\rangle|\lambda_{\perp}\rangle, |1\rangle|\lambda\rangle\}$ and let $|\lambda'\rangle$ be the projection of $|\sigma'\rangle$ on $\Lambda'$. Let $P'_{\lambda}(\sigma)$ be the probability of getting $|\lambda'\rangle$ state upon measuring $|\sigma'\rangle$. It is easy to check that 
\noindent   
\begin{equation}
|\sigma'\rangle = \sin \omega'|\lambda'\rangle + \cos \omega' |\lambda'^{\perp}\rangle,\ \ \ P'_{\lambda}(\sigma) = \sin^{2}\omega' = (9/10)P_{\lambda}(\sigma) + (1/10).  \label{lambdaprimedefine}
\end{equation}
Here $|\lambda'^{\perp}\rangle$ state is orthogonal to $|\lambda'\rangle$. We achieve the desired lower bound $P'_{\lambda}(\sigma) \geq 1/10$. Also,
\begin{equation}
P_{\lambda}(\sigma) \geq 4/9 \Longrightarrow P'_{\lambda}(\sigma) \geq 5/10,\ \ \ \ P_{\lambda}(\sigma) < 5/9 \Longrightarrow P'_{\lambda}(\sigma) < 6/10. \label{determination}
\end{equation}
Thus we need to determine if $P'_{\lambda}(\sigma) \geq 0.5$ or if $P'_{\lambda}(\sigma) < 0.6$.

	Consider the amplitude amplification operator $\mathcal{A} = I_{\sigma'}I_{\lambda'}$, where $I_{\sigma'}$ ($I_{\lambda'}$) denote the selective inversion of $|\sigma'\rangle$ ($|\lambda'\rangle$) state, i.e.
\noindent   
\begin{equation}
\mathcal{A} = I_{\sigma'}I_{\lambda'},\ \ I_{\sigma'} = \mathbbm{1}-2|\sigma'\rangle\langle \sigma'|,\ \ I_{\lambda'} = \mathbbm{1}-2|\lambda'\rangle\langle \lambda'|.  \label{primeamplitude}
\end{equation}
$I_{\lambda'}$ can be implemented by a selective inversion of $\Lambda'$ subspace spanned by the states $|0\rangle|\lambda\rangle, |0\rangle|\lambda^{\perp}\rangle$ and $|1\rangle|\lambda\rangle$. The states $\{|0\rangle|\lambda\rangle,|0\rangle|\lambda^{\perp}\rangle\}$ can be inverted by inverting the $|0\rangle$ state of ancilla qubit. The state $|1\rangle|\lambda\rangle$ is inverted by applying $I_{\lambda}$, the selective inversion of $|\lambda\rangle$ state, if and only if the ancilla qubit is in $|1\rangle$ state. In our case, we know the subspace corresponding to $|\lambda\rangle$ and hence $I_{\lambda}$ as well as $I_{\lambda'}$ are trivial to implement. To implement $I_{\sigma'}$, let $R$ be an operator such that $|\sigma'\rangle = R|0\rangle|\sigma\rangle = \left(1/\sqrt{10}\right)\left(|0\rangle + 3|1\rangle\right)|\sigma\rangle$. Then \noindent   
\begin{equation}
I_{\sigma'} = \mathbbm{1} - 2[R|0\rangle|\sigma\rangle\langle 0|\langle \sigma|R^{\dagger}] = RI_{0,\sigma}R^{\dagger}.
\end{equation}     	
Here $I_{0,\sigma} = \mathbbm{1} - 2|0\rangle|\sigma\rangle\langle 0|\langle \sigma|$ inverts the $|0\rangle|\sigma\rangle$ state and can be implemented by applying $I_{\sigma}$ if and only if the ancilla qubit is in $|0\rangle$ state. In our case, generally we don't know $|\sigma\rangle$ state hence the only non-trivial component of $\mathcal{A}$ is $I_{\sigma}$.

	The eigenspectrum of $\mathcal{A}$ has been analysed in detail in Section 2 of ~\cite{qaa2}. Its relevant eigenstates and the corresponding eigenvalues are  
\noindent   
\begin{equation}
\mathcal{A}|\Omega_{\pm}\rangle = e^{\pm i 2\omega'}|\Omega_{\pm}\rangle,\ \ \ \sqrt{2}|\Omega_{\pm}\rangle = |\lambda'\rangle \pm |\lambda'^{\perp}\rangle.
\end{equation}
In terms of these eigenstates, the state $|\sigma'\rangle$ is given by
\noindent   
\begin{equation}
\sqrt{2}|\sigma'\rangle = e^{i \omega'}|\Omega_{+}\rangle - e^{-i \omega'}|\Omega_{-}\rangle. \label{linearsuper}
\end{equation}

	Our algorithm $\mathcal{S}$ works with any state $|\kappa\rangle$ satisfying $|\langle \kappa|\Omega_{+}\rangle| = |\langle \kappa|\Omega_{-}\rangle| = 1/\sqrt{2}$. To this, we attach an ancilla qubit in the state $(1/\sqrt{2})(|0\rangle + |1\rangle)$. Thus the initial state is given by
\noindent
\begin{equation}
2|\kappa\rangle|+\rangle = e^{i \kappa_{+}}|\Omega_{+}\rangle|0\rangle + e^{i \kappa_{+}}|\Omega_{+}\rangle|1\rangle + e^{i \kappa_{-}}|\Omega_{-}\rangle|0\rangle + e^{i \kappa_{-}}|\Omega_{-}\rangle|1\rangle.
\end{equation}
We apply $\mathcal{A}$ on $|\kappa\rangle$ if the ancilla qubit is in $|1\rangle$ state. Upto a factor of $2$, we get
\noindent
\begin{equation}
e^{i \kappa_{+}}|\Omega_{+}\rangle\left(|0\rangle+e^{i 2\omega'}|1\rangle\right) + e^{i \kappa_{-}}|\Omega_{-}\rangle\left(|0\rangle+e^{-i 2\omega'}|1\rangle\right). 
\end{equation}   
We apply Hadamard gate $H$ on the ancilla qubit. Upto a factor of $\sqrt{2}$, we get
\noindent
\begin{equation}
\left( e^{i (\kappa_{+}+\omega')}|\Omega_{+}\rangle + e^{i (\kappa_{-}-\omega')}|\Omega_{-}\rangle \right)\cos \omega' |0\rangle - i\left( e^{i (\kappa_{+}+\omega')}|\Omega_{+}\rangle - e^{i (\kappa_{-}-\omega')}|\Omega_{-}\rangle \right)\sin \omega'|1\rangle. \label{afterHadamard} 
\end{equation}
We measure the ancilla qubit. The post-meaurement state is given by
\noindent   
\begin{equation}
\sqrt{2}|\kappa + 1\rangle = \left(e^{i (\kappa_{+}+\omega')}|\Omega_{+}\rangle + (-1)^{X_{\kappa}} e^{i (\kappa_{-}-\omega')}|\Omega_{-}\rangle\right)|X_{\kappa}\rangle.  \label{postmeasure}
\end{equation}
\begin{equation}
(\kappa+1)_{+} = \kappa_{+}+\omega',\ \ \ (\kappa+1)_{-} = \kappa_{-}-\omega' + \pi X_{\kappa}. \label{kappaplus}
\end{equation}
Here $X_{\kappa}$ denote the measurement outcome with probabilities given by Eq. (\ref{afterHadamard}) as
\noindent   
\begin{equation}
\rm Prob(X_{\kappa} = 1) = \sin^{2}\omega' = P'_{\lambda}(\sigma),\ \ \ \rm Prob(X_{\kappa} = 0) = 1- P'_{\lambda}(\sigma). \label{measurementprobabilities}
\end{equation}

	We iterate the above process $q$ times to get the state $|\kappa + q\rangle$. Eq. (\ref{kappaplus}) implies that
\noindent   
\begin{equation}
(\kappa + q)_{+} = \kappa_{+} + q\omega',\ \ \ (\kappa + q)_{-} = \kappa_{-} - q\omega' + \pi (N_{1} \rm mod 2),\ \ \ N_{1} = \sum_{\kappa}^{\kappa + q-1}X_{\kappa}\ .  \label{kappaplusq}
\end{equation}
We have used $\rm mod 2$ here as the angles $\kappa_{\pm}$ are same modulo $2\pi$. The definition $N_{1} = \sum_{\kappa}^{\kappa + q-1}X_{\kappa}$ implies that $N_{1}$ is the total number of $1$'s that we get as measurement outcomes during $q$ iterations. In a single iteration, the probability of getting $1$ after measurement is $\rm Prob(X_{\kappa} =1) = P'_{\lambda}(\sigma)$ due to Eq. (\ref{measurementprobabilities}). Hence the probability distribution of $N_{1}$ is a binomial distribution having a sharp peak at $N_{1}^{\rm max} = qP'_{\lambda}(\sigma)$ and decaying exponentially as we go away from this peak. This exponential decay can be quantified using Hoeffding's bound~\cite{hoeffding} which states that
\noindent   
\begin{equation}
\rm Prob\left(|N_{1}-N_{1}^{\rm max}|/q > t\right) = \rm Prob\left(\left|(N_{1}/q)-P'_{\lambda}(\sigma)\right| > t\right) \leq e^{-2qt^{2}}.
\end{equation}
Thus if $N_{1}/q \geq 0.55$ then $\rm Prob(P'_{\lambda}(\sigma) < 0.5)$ is at most $e^{-q/200}$ and hence $\rm Prob(P'_{\lambda}(\sigma) \geq 0.5)$ is at least $1-e^{-q/200}$. Similarly if $N_{1}/q < 0.55$ then $\rm Prob(P'_{\lambda}(\sigma)< 0.6)$ is at least $1-e^{-q/200}$. Thus,
\noindent
\begin{equation}
N_{1} \geq 0.55q \Longrightarrow P'_{\lambda}(\sigma)\geq 0.5,\ \ \ \ N_{1} < 0.55q \Longrightarrow P'_{\lambda}(\sigma) < 0.6, \label{N1P}
\end{equation}
with the error probability $e^{-q/200}$. This performs the desired determination given by Eq. (\ref{determination}).

	A good thing about the algorithm $\mathcal{S}$ is that we can get back the initial state $|\kappa\rangle$. We choose $q$ to be even. If $N_{1}$ is also even then Eq. (\ref{kappaplusq}) and $\mathcal{A}|\Omega_{\pm}\rangle = e^{\pm i 2\omega'}|\Omega_{\pm}\rangle$ imply that
\noindent   
\begin{equation}
|\kappa + q\rangle =  \mathcal{A}^{q/2}|\kappa\rangle \Longrightarrow |\kappa\rangle = (\mathcal{A}^{\dagger})^{q/2}|\kappa+ q\rangle. \label{sigmastateobtain}  
\end{equation}   
Thus $q/2$ applications of $\mathcal{A}$ on $|\kappa + q\rangle$ brings back the initial state $|\kappa\rangle$. If $N_{1}$ is odd, we add $q_{e} = 2$ extra iterations. It keeps $q_{\rm tot} = q + q_{e}$, the total number of iterations, even but $N_{1}$ remains odd only if we get $\{0,0\}$ or $\{1,1\}$ as the $2$ measurement outcomes in $2$ extra iterations. As the probability of getting $1$ in single measurement is $P'_{\lambda}(\sigma)$, the probability of $N_{1}$ remaining odd is $(1-P'_{\lambda}(\sigma))^{2}+(P'_{\lambda}(\sigma))^{2} = 1-2P'_{\lambda}(\sigma)[1-P'_{\lambda}(\sigma)]$. We keep on adding $2$ extra iterations till $N_{1}$ becomes even. The error probability of $N_{1}$ remaining odd after $q_{e}$ extra iterations is $[1-2P'_{\lambda}(\sigma)\{1-P'_{\lambda}(\sigma)\}]^{q_{e}/2}$. As $P'_{\lambda}(\sigma) \geq 1/10$, the error probability is at most $0.91^{q_{e}} < e^{-0.09q_{e}}$. Note that in the absence of lower bound on $P'_{\lambda}(\sigma)$, we may have to perform large number of extra iterations to get back the initial state. That makes the algorithm slightly inconvenient even though expected number of iterations is not much large. 

Choosing $q_{e} = q/10$, the error term $e^{-0.09q_{e}}$ becomes negligible compared to the error term $e^{-q/200}$ in Eq. (\ref{N1P}). Hence $q_{\rm tot} = q+q_{e} =  1.1q$ iterations reduce the error probability to $e^{-q/200} = e^{-q_{\rm tot}/220}$. If the desirable error probability is $\epsilon_{1}$ then 
\noindent   
\begin{equation}
\epsilon_{1} = e^{-q/220} \Longrightarrow q \geq 220 \ln (1/\epsilon_{1}) = \Theta(\ln(1/\epsilon_{1})).
\end{equation}
Here we have omitted the subscript $\rm tot$ from $q_{\rm tot}$. We can use the $|\sigma\rangle$ state as our initial state $|\kappa\rangle$ as Eq. (\ref{linearsuper}) implies that $|\sigma\rangle$ satisfies the condition $|\langle \sigma | \Omega_{+}\rangle| = |\langle \sigma |\Omega_{-}\rangle| = 1/\sqrt{2}$ required for $|\kappa\rangle$ state. Thus $\mathcal{S}$ uses $3q/2 = \Theta(\ln (1/\epsilon_{1}))$ applications of $\mathcal{A}$: $q$ applications for $q$ iterations and $q/2$ to get back the $|\sigma\rangle$ state as implied by Eq. (\ref{sigmastateobtain}). 

	Consider the algorithm $\mathcal{S}_{g\gamma}$ to find $P_{g\gamma}$ where $P_{0\gamma} = P_{12}^{27}(\psi,f_{\gamma})$ and $P_{1\gamma} = P_{20}^{3}(\psi,f_{\gamma})$. We choose $|\sigma\rangle = |\psi,f_{\gamma}\rangle$ and $\Lambda$ to be the subspace $\Lambda_{12}^{27}$ spanned by $k \in [12,27]$ (for $P_{0\gamma}$) or the subspace $\Lambda_{20}^{3}$ spanned by $k \in [20,3]$ for $P_{1\gamma}$. The algorithm $\mathcal{S}_{g\gamma}$ uses $\Theta(\ln(1/\epsilon_{1}))$ applications of $\mathcal{A}_{g\gamma} = I_{\psi,f_{\gamma}'}I_{\lambda'}$. As discussed after Eq. (\ref{primeamplitude}), $\mathcal{A}_{g\gamma}$ can be implemented if we can implement $I_{\psi,f_{\gamma}}$ and $I_{\lambda}$. Implementation of $I_{\lambda}$ is easy as we know the corresponding subspace $\Lambda$. To implement $I_{\psi,f_{\gamma}}$, note that $|\psi,f_{\gamma}\rangle = \mathcal{E}(U_{\gamma})|\psi,0_{5}\rangle$. Hence
\noindent   
\begin{equation}
I_{\psi,f} = \mathbbm{1} - \mathcal{E}(U_{\gamma})I_{\psi,0_{5}}\mathcal{E}(U_{\gamma})^{\dagger},\ \ \ I_{\psi,0_{5}} = \mathbbm{1} - 2|\psi,0_{5}\rangle\langle \psi,0_{5}|.
\end{equation} 
The operator $I_{\psi,0_{5}}$ is the selective inversion of $|\psi,0_{5}\rangle$ state and can be implemented by applying $I_{\psi}$, the selective inversion of $|\psi\rangle$ state, if and only if all $5$ ancilla qubits are in $|0\rangle$ state. Also $\mathcal{E}(U_{\gamma})$ uses $2^{5} = 32$ applications of $U_{\gamma}$. Thus implementing $\mathcal{A}_{g\gamma}$ needs $64$ applications of $U_{\gamma} = U^{2^{\gamma}}$ and $1$ application of $I_{\psi}$. Hence $\mathcal{S}_{g\gamma}$ needs $\Theta(2^{\gamma}\ln (1/\epsilon_{1}))$ applications of $U$ and $\Theta(\ln (1/\epsilon_{1}))$ applications of $I_{\psi}$.

	To use the algorithm $\mathcal{S}_{g\gamma}$ for phase estimation, Eq. (\ref{numberforphase}) implies that to estimate $\phi$ to an accuracy of $\delta = 2^{-\Gamma}$, we need a total of $4[\log_{2}(1/\delta)+1] \approx 4\log_{2}(1/\delta)$ applications of $\mathcal{S}_{g\gamma}$ with $4$ applications for each value of $\gamma \in \{0,1,\ldots,\Gamma\}$. Suppose the desired error probability in phase estimation is $\epsilon_{2}$. Then the error probability $\epsilon_{1}$ in implementing each $\mathcal{S}_{g\gamma}$ must satisfy 
\begin{equation}
\epsilon_{1} = \frac{\epsilon_{2}}{4\log_{2}(1/\delta)}.  \label{epsilon12}
\end{equation}
As $\mathcal{S}_{g\gamma}$ uses $\Theta(2^{\gamma}\ln(1/\epsilon_{1}))$ applications of $U$, the total number of applications used by our phase estimation algorithm is
\begin{equation}
Q(U,\epsilon_{2}) = 4\sum_{\gamma = 0}^{\Gamma}\Theta\left(2^{\gamma}\ln\left(\frac{1}{\epsilon_{1}}\right)\right) = \Theta\left[\frac{1}{\delta}\ln\left(\frac{\ln(1/\delta)}{\epsilon_{2}}\right)\right], \label{QU}
\end{equation}
where we have made use of Eq. (\ref{epsilon12}). Similarly, the number of applications of $I_{\psi}$ used by our algorithm is
\begin{equation}
Q(I_{\psi},\epsilon_{2}) = 4\sum_{\gamma = 0}^{\Gamma}\Theta\left(\ln\left(\frac{1}{\epsilon_{1}}\right)\right) = \Theta\left[\ln\left(\frac{1}{\delta}\right)\ln\left(\frac{\ln(1/\delta)}{\epsilon_{2}}\right)\right]. \label{QIpsi}
\end{equation}
Above two equations determine the time complexity of our phase estimation algorithm that uses an approximate eigenstate $|\psi\rangle$ to estimate $\phi$ to an accuracy of $\delta$ with the success probability $1-\epsilon_{2}$.
	
\section{Application to Eigenpath Traversal Problem}
\noindent

Let $|\theta_{s}\rangle$ be the non-degenerate eigenstates of unitary operators $V_{s}$ with the eigenvalues $e^{i 2\pi \theta_{s}}$ for $s \in \{1,2,\ldots,n\}$. It is promised that for any $s$, the eigenphase $\theta_{s}$ has a minimum spectral gap of $\Delta$ from non-$\theta_{s}$ eigenphases of $V_{s}$. In the eigenpath traversal problem, the goal is to evolve a given copy of $|\theta_{1}\rangle$ to $|\theta_{n}\rangle$. For this problem, the algorithm with optimal time complexity (upto a logarithmic factor) is the BKS algorithm due to Boixo, Knill and Somma~\cite{BKS}. This algorithm is basically a series of $n-1$ transformations from $|\theta_{s}\rangle$ to $|\theta_{s+1}\rangle$ for $s \in \{1,2,\ldots,n-1\}$, starting with $s =1$. Assuming $|\langle \theta_{s}|\theta_{s+1}\rangle|^{2} \geq 1/3$, the transformation $|\theta_{s}\rangle \rightarrow |\theta_{s+1}\rangle$ is done using fixed-point quantum search algorithms~\cite{fpqs1,fpqs2} which involves $\Theta(1)$ implementations of the selective inversions $I_{\theta_{s'}} = \mathbbm{1} - 2|\theta_{s'}\rangle\langle \theta_{s'}|$ for $s' = s$ and $s' = s+1$ (see Theorem V.1 of ~\cite{BKS}). Thus BKS algorithm requires $\Theta(n)$ implementations of $I_{\theta_{s}}$.

	The operator $I_{\theta_{s}}$ is implemented using the knowledge of $\theta_{s}$ to an accuracy of $\Delta/4$. This implementation was presented in ~\cite{highPEA} and used as a reflection oracle in the BKS algorithm (Definition III.3 of ~\cite{BKS}). A detailed implementation scheme was recently presented in Section III of ~\cite{postprocessing}. We need $O(\ln (1/\epsilon_{3})/\Delta)$ applications of $V_{s}$ to implement $I_{\theta_{s}}$ where $\epsilon_{3}$ is the desired failure probability. As BKS algorithm uses $\Theta(n)$ implementations of $I_{\theta_{s}}$, for each implementation, we must have $\epsilon_{3} = \epsilon/\Theta(n)$ if $\epsilon$ is the desired error probability of BKS algorithm. Hence the number of applications of $V_{s}$ required by BKS algorithm is
\begin{equation}
\mathcal{Q}_{1} = \Theta\left(\frac{n}{\Delta}\ln\frac{n}{\epsilon}\right). \label{mathcalN}
\end{equation}

	What if we don't have a prior knowledge of $\theta_{s}$? How do we implement $I_{\theta_{s}}$ then. Suppose we have a copy of $|\theta_{s}\rangle$. Then PEA can be used to estimate $\theta_{s}$ to an accuracy of $O(\Delta)$ using $O(1/\Delta)$ applications of $V_{s}$ and then $I_{\theta_{s}}$ can be implemented. But we also need to implement $I_{\theta_{s+1}}$ to transform $|\theta_{s}\rangle$ to $|\theta_{s+1}\rangle$. How do we estimate $\theta_{s+1}$ using the copy of $\theta_{s}$? To do so, BKS algorithm assumes $|\langle \theta_{s}|\theta_{s+1}\rangle|^{2} \geq 2/3$ for all $s$ (see Theorem V.2 of ~\cite{BKS}) so that $|\theta_{s}\rangle$ serves as an approximate eigenstate for the operator $V_{s+1}$. Then BKS algorithm uses the method of multiple copies described in Section $1$ to estimate $\theta_{s+1}$. Thus it successively transforms $|\theta_{s}\rangle^{\otimes r}$ to $|\theta_{s+1}\rangle^{\otimes r}$ for $s \in \{1,2,\ldots,n-1\}$, starting with $s=1$. We need $r = \Theta (\ln n)$ copies of $|\theta_{1}\rangle$ to achieve a constant success probability. Using $r$ copies of $|\theta_{1}\rangle$ increases the spatial complexity of BKS algorithm by a factor of $r = \Theta(\ln n)$. To prevent this, we use the phase estimation algorithm with approximate eigenstate ($\rm PEA_{\approx}$) described in this paper. It requires only one copy of $|\theta_{1}\rangle$. 

	To find the required number of applications of $V_{s} \equiv U$, we put $\delta = \Delta/4$ in Eqs. (\ref{QU},\ref{QIpsi}) as that is the desired accuracy of estimation of $\theta_{s}$. Also, the desired error probability of BKS algorithm is $\epsilon$ and as BKS algorithm uses $n$ instances of $\rm PEA_{\approx}$, we put $\epsilon_{2} = \epsilon/n$. Each $I_{\theta_{s}}$ can be implemented using $O(\ln (1/\epsilon_{3})/\Delta)$ applications of $V_{s}$. A single instance of $\rm PEA_{\approx}$ uses $Q(I_{\psi},\epsilon_{2})$ applications of $I_{\psi}$ and hence BKS algorithm uses $nQ(I_{\psi},\epsilon_{2})$ applications of $I_{\psi}$. Thus $\epsilon_{3} = \epsilon/nQ(I_{\psi},\epsilon_{2})$. With all these values, the total required number of applications of $V_{s}$ is
\begin{equation}
\mathcal{Q} = nQ\left(U,\frac{\epsilon}{n}\right) + Q\left(I_{\psi},\frac{\epsilon}{n}\right)\frac{n}{\Delta}\ln\frac{nQ(I_{\psi},\epsilon/n)}{\epsilon} 
\end{equation} 
Eqs. (\ref{QU}) and (\ref{QIpsi}) also imply that $Q(I_{\psi},\epsilon_{2}) = Q(U,\epsilon_{2})\Theta(\Delta\ln (1/\Delta))$. Using this and after little calculation, above equation reduces to 
\begin{equation}
\mathcal{Q} = \Theta\left(\frac{n}{\Delta}\ln\frac{1}{\Delta}\right) \ln \beta \ln (\beta\ln \beta),\ \ \  \beta = \frac{n}{\epsilon}\ln\frac{1}{\Delta}\ .
\end{equation}
We need $\mathcal{Q}$ applications of $V_{s}$ to estimate all $\theta_{s}$ to the desired accuracy of $\Delta/4$. After this estimation, BKS algorithm can transform $|\theta_{1}\rangle$ to $|\theta_{n}\rangle$ using $\mathcal{Q}_{1}$ applications of $V_{s}$, where $\mathcal{Q}_{1}$ is given by Eq. (\ref{mathcalN}). $\mathcal{Q}$ is larger than $\mathcal{Q}_{1}$ only by a logarithmic factor. Thus the time complexity of BKS algorithm increases only by a logarithmic factor. Whereas the spatial complexity is improved tremendously from $\Theta(\ln n)$ copies to just $1$ copy of eigenstate.

	As shown in Lemma V.3 of ~\cite{BKS}, once we have found $\theta_{s}$ for all $s$ to the desired accuracy of $\Delta/4$, another sequence of eigenstates $|\theta'_{s}\rangle$ can be found such that $|\theta'_{1}\rangle = |\theta_{1}\rangle$ and $|\theta'_{n'}\rangle = |\theta_{n}\rangle$, where $n'  = O(L)\leq n$. The quantity $L$ is the \emph{angular path length} defined by $L = \sup\left(\sum_{s=1}^{n}\cos^{-1}|\langle\theta_{s}|\theta_{s-1}\rangle|\right)$. For continuous paths, if $|\theta_{s}\rangle$ is differentiable in $s$, an alternative expression is $L = \int_{0}^{1}ds\||(\partial \theta_{s}/\partial s)\rangle\|$. Thus BKS algorithm has a time complexity of $\Theta(L/\Delta)$ ignoring logarithmic factors. It is better than the earlier algorithm with complexity $\Theta(L^{2}/\Delta)$ based on quantum zeno effect and phase randomization~\cite{randomization}. Furthermore, the time complexity of $\Theta(L/\Delta)$ has also been proved to be optimal~\cite{BKSoptimal}. 

	A special case of eigenpath traversal problem is the quantum adiabatic evolution where $V_{s} = exp(-i H_{\hat{s}}t)$, where $\hat{s} = (s-1)/(n-1)$ and $H_{\hat{s}} = (1-\hat{s})H_{0} + \hat{s}H_{1}$ is the interpolating Hamiltonian between $H_{0}$ and $H_{1}$. It has applications to quantum computation~\cite{farhi}. In this case, $L$ is $O(\|H_{0}+H_{1}\|/\Delta)$. Childs et.al. presented an algorithm based on quantum zeno effect to simulate adiabatic evolution~\cite{childs} using $O(L^{2}/\Delta) = O(1/\Delta^{3})$ time steps. BKS algorithm can do this using $O(L/\Delta) = O(1/\Delta^{2})$ time steps. As shown in ~\cite{digital}, for adiabatic evolution, we don't need phase estimation with approximate eigenstates as perturbation theory gives sufficient estimates of $\theta_{s}$.  The time complexity $O(1/\Delta^{2})$ of BKS algorithm is same as the evolution time $O(1/\Delta^{2})$ required by folk adiabatic approximation~\cite{adiabatic} and better than the evolution time $O(1/\Delta^{3})$ required by rigorous adiabatic approximation~\cite{adiabatic1,adiabatic2,adiabatic3}. Note that the operators $V_{s} = e^{-i H_{\hat{s}}t}$ can be efficiently simulated for sparse Hamiltonians using recently developed simulation algorithms~\cite{berry}. 

	Recently, it was shown that if the adiabatic evolution involves only the ground state then $L = O(1/\Delta^{1/2})$, much less than $O(1/\Delta)$ for $\Delta \ll 1$~\cite{PRAimprovedtime}. Then BKS algorithm has the time complexity $\Theta(1/\Delta^{3/2})$. But unlike the case of $L = O(1/\Delta)$, perturbation theory fails to give sufficient estimates of $\theta_{s}$ in this case. Hence, using our phase estimation algorithm with approximate eigenstates, we can use BKS algorithm in this case with just a single copy of the initial eigenstate.

\section{Discussion and Conclusion}
\noindent

	We have assumed $|\langle \psi |\phi\rangle|^{2} \geq 2/3$ only for the simplicity of presentation. It also matches the lower bound assumed by BKS algorithm. Similar ideas can be used if $|\langle \psi|\phi\rangle|^{2} \geq (1/2)+h$, where $h > 0$ is any small constant. Rather than choosing $c = 4$ in Eq. (\ref{probabilityeqn}), we choose $c$ such that $1/2(c-1) = h$, i.e. $c \approx 1/2h$. Then, in place of $5/9$, the lower bound in Eq. (\ref{eqnc4}) becomes $(\frac{1}{2}+h)(1-h) = \frac{1}{2}+\frac{h}{2}$. Similar ideas can be used to handle this case. We just need to increase the number of ancilla qubits used in PEA from $\mu = 5$ to $\mu = 6+\log_{2}h$ so that the Hilbert space of ancilla qubits has dimension $128c$. Using more number of ancilla qubits also increases exponentially the required number of applications of $U$. Also, while using the probability estimation algorithm $\mathcal{S}_{g\gamma}$, in place of determing if $P_{g\gamma} \geq 4/9$ or $P_{g\gamma} < 5/9$, we need to determine if $P_{g\gamma} \geq (1-h)/2$ or $P_{g\gamma} < (1+h)/2$. It increases the number of iterations $q$ required by the algorithm $\mathcal{S}_{g\gamma}$ as the error probability decreases as $e^{-qh^{2}/2}$ which can be much larger than $e^{-q/220}$ for small $h$. To compensate for it, we must choose suitably large values of $q$. The details can be worked out easily.

	We also point out that in general, our algorithm cannot be used to detect if $|\langle \psi|\phi\rangle|^{2} \geq 2/3$ or not. It just assumes $|\langle \psi|\phi\rangle|^{2} \geq 2/3$ and then outputs the corresponding value of $\phi$. Generally our algorithm cannot determine if this assumption is true or not. Basically, our algorithm shows that the assumption $|\langle \psi|\phi\rangle|^{2} \geq 2/3$ and the ability to implement $I_{\psi}$ is enough to apply a selective inversion $I_{\phi}$ of $|\phi\rangle$ state by finding the corresponding eigenphase $\phi$.

	Our algorithm may find applications in finding the ground state energies of quantum systems like molecules. It is not necessary to have a perfect copy of the ground state. Any state with sufficiently high overlap can be used provided we can implement its selective inversion. It is possible if such an approximate eigenstate is the eigenstate of some known perturbed Hamiltonian with a sufficiently known eigenvalue. We believe that our algorithm can also find other important applications in quantum computing.

\nonumsection{References}

\end{document}